\documentclass[a4paper,12pt]{article}
\usepackage{amsmath}
\usepackage{amssymb}
\usepackage{latexsym}
\usepackage{lscape}
\usepackage{pdflscape}
\usepackage{hyperref}
\usepackage[dvipsnames]{xcolor}
\usepackage{amsfonts}
\usepackage{amsthm}
\usepackage{latexsym}
\usepackage{enumitem}
\usepackage{graphics}
\usepackage{bbm}
\usepackage{tensor}
\usepackage[framemethod=tikz,skipabove=1pt,innertopmargin = 0pt,skipbelow = 1pt,innerbottommargin = 0pt,innerleftmargin = 0pt,innerrightmargin = 0pt,leftmargin = 0pt,rightmargin = 0pt]{mdframed} 
\usepackage{hyperref}
\newcommand{\be}{\begin{equation}}
\newcommand{\ee}{\end{equation}}
\newcommand{\ben}{\begin{equation*}}
\newcommand{\een}{\end{equation*}}
\newcommand{\bean}{\begin{eqnarray*}}
\newcommand{\eean}{\end{eqnarray*}}
\def\bal#1\eal{\begin{align}#1\end{align}}
\newcommand{\ph}{\phantom}
\newcommand{\bsub}{\begin{subequations}}
\newcommand{\esub}{\end{subequations}}

\newcommand{\disfrac}[1][2]{\displaystyle\frac}

\newcommand{\non}{\nonumber}

\newcommand{\ima}{\mathbbmtt{i}} 
\newcommand{\drm}{\textrm{d}}
\newcommand*\xbar[1]{%
  \hbox{%
    \vbox{%
      \hrule height 0.5pt 
      \kern0.5ex
      \hbox{%
        \kern-0.1em
        \ensuremath{#1}%
        \kern-0.1em
      }%
    }%
  }%
}
\newtheorem{theorem}{Theorem}

\numberwithin{equation}{section}
\begin{document}

\title{\textbf{Variational Contact Symmetries of
Constraint Lagrangians}}
\vspace{1cm}
\author{\textbf{Petros A. Terzis}$^a$\thanks{pterzis@phys.uoa.gr},\, \textbf{N. Dimakis}$^b$\thanks{nsdimakis@gmail.com}\,, \textbf{T. Christodoulakis}$^a$\thanks{tchris@phys.uoa.gr},\\ \textbf{Andronikos Paliathanasis} $^{c,d}$\thanks{paliathanasis@na.infn.it},\, \textbf{Michael Tsamparlis$^e$}\thanks{mtsampa@phys.uoa.gr}\\
{$^a$\it Nuclear and Particle Physics Section, Physics
Department,}\\
{\it University of Athens, GR 157--71 Athens}\\
{$^b$ \it Instituto de Ciencias Fisicas y Matematicas,}\\
{\it Universidad Austral de Chile, Valdivia, Chile}\\
{$^c$ \it Dipartimento di Fisica, Universit\`{a} di Napoli ``Federico
II''}\\
{\it Complesso universitario Monte S.~Angelo, Via, Cintia 9, I- 80126 Napoli, Italy}\\
{$^d$ \it Istituto Nazionale di Fisica Nucleare (INFN) sezione di Napoli}\\
{\it Complesso universitario Monte S.~Angelo, Via Cintia 9, I- 80126 Napoli,
Italy}\\
{$^e$ \it Department of Physics, Section of Astronomy, Astrophysics and
Mechanics,}\\
{\it University of Athens, Panepistemiopolis, Athens 157 83, Greece}
}
\date{}
\maketitle
\vspace{-1cm}

\abstract{The investigation of contact symmetries of re--parametrization
invariant Lagrangians of finite degrees of freedom and quadratic in the velocities is presented.
The main concern of the paper is those symmetry generators which
depend linearly in the velocities. A natural extension of the
symmetry generator along the lapse function $N(t)$, with the appropriate
extension of the dependence in $\dot{N}(t)$ of the gauge function,
is assumed; this action yields new results.
The central finding is that the integrals of motion are either
linear or quadratic in velocities and are generated,
respectively by the conformal Killing vector fields and the
conformal Killing tensors of the configuration space  metric deduced from the kinetic part of the Lagrangian (with appropriate conformal factors).
The freedom of re--parametrization allows one to appropriately scale $N(t)$, so that the
potential becomes constant; in this case the integrals of motion can be constructed
from the Killing fields and Killing tensors of the scaled metric.
A rather interesting result is the non--necessity of the gauge function
in Noether's theorem due to the presence of the Hamiltonian constraint.}

\section{Introduction}
The Lie symmetry method is a systematic tool for the study of systems of differential equations.
The importance of Lie symmetries lies in the fact that they
can be used in order to reduce the order of an ordinary differential equation or to assist in the solution of differential systems
by reducing the number of independent variables.
The Lie symmetries which leave invariant a variational integral are called Noether symmetries.
Noether symmetries form a subalgebra of the Lie symmetries of the equations following
from the variational integral. The characteristic of Noether symmetries is that to each symmetry there corresponds a
divergence free current along with the corresponding charge.
The well known conservation laws of energy, angular momentum all follow from Noether symmetries.
It is well known that conservation laws are important tools which can be used
for the determination of integrable manifolds
of a dynamical system in mathematical physics,
biology, economics and many others \cite{Dask06,Camci14,IgorL,Zhdanov,Huang,Azad07,Mustafa2015,Ibrag98,Nucci2005}.

The determination of the Lie and Noether symmetries of a given differential equation consists of two steps,
(a) The derivation of the symmetry conditions and (b) the solution of these conditions. The first step is formal and it is
outlined in e.g. \cite{Bluman,Stephani,OlverODE}. The second step can be a difficult task since
the symmetry conditions can be quite involved. Tsamparlis \& Paliathanasis in \cite{Tsamparlis:Geod,MT2d} proposed
a geometric method for the solution of the symmetry conditions for
regular Hamiltonian systems which describe the motion of a particle in a Riemannian space under the action of
a potential. It was shown that the Lie and Noether
symmetries of that system are related with the special
projective algebra of the underlying space.
This geometric approach has been extended to the determination
of the Lie and Noether point symmetries
of some families of partial differential equations \cite{APpde,ApKG}.

A similar analysis has been established by Christodoulakis, Dimakis \& Terzis \cite{CDT2014} for constraint Lagrangians quadratic in the velocities, where it was shown that the point symmetries of equations of motion are exactly the variational symmetries (containing the time reparametrization symmetry) plus the scaling symmetry. The variational symmetries was seen to be the simultaneous conformal
Killing vector fields of both the metric, defined by the kinetic term, and the potential.

These results have been applied in various areas, i.e. in classical mechanics,
in general relativity and in cosmology; in each case new dynamical
systems which admit symmetries were found for instance see
\cite{Terzis:2014cra,Christodoulakis:2013sya, Christodoulakis:2012eg,MT2d,MT3d,Christodoulakis:2014wba,Dimakis:2013oza,Basilakos:2011rx,Ap2sf} and reference therein.

In this work we extend this geometric approach in order to study the Lie B\"{a}cklund
variational symmetries (or dynamical Noether symmetries) of constrained Lagrangian systems. Specifically, we study the symmetries
which arise from contact transformations, the contact Noether symmetries. Contrary to the point transformations
which are defined in the configuration space manifold, where the dynamical equations are defined,
the contact transformations are defined in the tangent bundle thus they depend also in the velocities \cite{Katzin,RoyMartT}.
Such contact Noether symmetries of regular Lagrangian systems with a potential have been studied in \cite{Kalotas:Noether} and
it was shown that the corresponding conservation laws follow from the Killing tensors of the kinetic metric and the potential.
Some important conservation laws of that kind in physics are: the Runge-Lenz vector field of the Kepler problem,
the Ray-Reid invariant, and the Carter constant in Kerr spacetime \cite{Ballesteros09,OConnell,Reid_Ray,Carter}.
 As it will be shown, in the case of constrained Lagrangian systems the results are different from that of \cite{Kalotas:Noether}
 in that the Noether contact symmetries are related to the Conformal Killing tensors of the underlying space.
The method we develop can be used and for regular Lagrangian systems. Furthermore it is interesting to note
that, in the presence of constrains, the gauge function of Noether's theorem does not play a significant role:
the presence of the quadratic constraint renders it non--essential in finding the conservation laws.
The plan of the paper is as follows.

In section \ref{preliminaries}, we give the basic theory of the Lie B\"{a}cklund symmetries.
In section \ref{section3}, we study the linear contact symmetries of constraint Lagrangian systems and we prove
the main theorem of this work. The conservation laws which follow from the contact symmetries are studied in
\ref{section4}. Furthermore, in section \ref{examples} we demonstrate the main results in two applications, (a)
the determination of the variational contact symmetries for the geodesic equations of a B\textit{iv}
pp-wave spacetime and (b) the derivation of the Ray-Reid invariant of corresponding constraint Hamiltonian system.
Finally, in section \ref{disc} we discuss our results.

\section{Preliminaries}
\label{preliminaries}

For the convenience of the reader, in this section we present
the basic properties and definitions concerning the generalized symmetries.

Let $H=H\left( x^{i},u^{A},u_{,i}^{A},u_{,ij}^{A}...\right) $ be a function
which is defined in the space $A_{M}=\left\{
x^{i},u^{A},u_{,i}^{A},u_{,ij}^{A},...\right\} $ where $x^{i}$ are $n$
independent variables and $u^{A}$ are $m$ dependent variables and $%
u_{,i}^{A}=\frac{\partial u^{A}}{\partial x^{i}}$. The function $H$ describes a
set of differential equations, for $n=1$ of ordinary differential equatons
and for $n>1$ a set of partial differential equations.

We shall say that the function $H$ will be invariant under the action of the
following infinitesimal transformation
\begin{align}
\bar{x}^{i}& =x^{i}+\varepsilon \xi ^{i}\left(
x^{i},u^{B},u_{,i}^{B},u_{,ij}^{B}...\right)   \label{LB.01} \\
\bar{u}^{A}& =u^{A}+\varepsilon \eta ^{A}\left(
x^{i},u^{B},u_{,i}^{B},u_{,ij}^{B}...\right)   \label{LB.02}
\end{align}%
if there exist a function $\lambda \left( x^{i},u,u_{,i},u_{,ij}...\right) $
such as the following condition holds \cite{Bluman}%
\begin{equation}
\left[ X,H\right] =\lambda H~\ ,~\mathrm{mod}H=0.  \label{LB.02a}
\end{equation}%
where $X=\frac{\partial \bar{x}}{\partial \varepsilon }\partial _{i}+\frac{%
\partial \bar{u}^{A}}{\partial \varepsilon }\partial _{A}$ is the generator
of the infinitesimal transformation (\ref{LB.01}), (\ref{LB.02}), i.e.
\begin{equation}
X=\xi ^{i}\left( x^{i},u^{B},u_{,i}^{B},u_{,ij}^{B}...\right) \partial
_{i}+\eta ^{A}\left( x^{i},u^{B},u_{,i}^{B},u_{,ij}^{B}...\right) \partial
_{u}  \label{LB.03}
\end{equation}%
and $\left[ .,.\right] $ is the Lie Bracket. When condition (\ref{LB.02a})
holds, the vector field $X$ is called Lie B\"{a}cklund symmetry of the
function $H$;  therefore we conclude that a Lie B\"{a}cklund transformation
is a transformation in $A_{M}$ which preserves the set of solutions $u^{A}$
of $H\left( x^{i},u,u_{,i},u_{,ij}...\right) $ in $A_{M}$. An alternative
way to write condition (\ref{LB.02a}) is%
\begin{equation}
X^{\left[ n\right] }H=\lambda H~\ ,~\mathrm{mod}H=0
\end{equation}%
where $X^{\left[ n\right] }$ is the $n$th prologation of $X$ in the space $%
A_{M}.$

Every differential equation admits as  Lie B\"{a}cklund symmetry the vector
field $D_{i}=\partial _{i}+u_{,i}\partial _{u}+u_{,ij}\partial _{u_{i}}+...$
which is the total derivative operator \cite{Stephani}. \ Let $X$ \ be a Lie B%
\"{a}cklund symmetry then the vector field
\begin{equation}
\bar{X}=X-f^{i}D_{i}=\left( \xi ^{k}-f^{k}\right) \partial _{k}+\left( \eta
^{A}-f^{k}u_{,k}^{A}\right) \partial _{u^{A}}+...
\end{equation}%
is also a Lie B\"{a}cklund symmetry for arbitrary functions $f^{i}$. Without
loss of generality we could select $f^{i}=\xi ^{i}$ and obtain%
\begin{equation}
\bar{X}=\left( \eta ^{A}-\xi ^{k}u_{,k}^{A}\right) \partial _{u^{A}}.
\label{LB.04}
\end{equation}

The generator (\ref{LB.04}) is called the canonical form of (\ref{LB.03}).
Since we can always absorb the term $\xi ^{k}u_{k}$ inside the $\eta $ we
conclude that $\bar{X}=Z^{A}\left(
x^{i},u^{B},u_{,i}^{B},u_{,ij}^{B}...\right) \partial _{u^{A}}$ is the
generator of a Lie B\"{a}cklund symmetry. This form of the generator is also suitable for the variational symmetries of the action e.g. see \cite{OlverODE} p. 331--333.

A special class of Lie B\"{a}cklund symmetries are the contact symmetries defined by the requirement that
the generator depends only on the first derivatives $u_{,i}$, i.e. it has
the general canonical form%
\begin{equation}
X_{C}=Z^{A}\left( x^{i},u^{B},u_{,i}^{B}\right) \partial _{u^{A}}.
\label{LB.04a}
\end{equation}

\subsection{Noether's Theorem for contact transformations}

Consider now that $H=H\left( t,q^{A},\dot{q}^{A},\ddot{q}^{A}\right) ~$where $\dot{q}^{A}=%
\frac{dq^{A}}{dt},$ and $H$ follows from a variational principle, that is
there exist a Lagrange function $L\left( t,q^{A},\dot{q}^{A}\right) $ such
as $\mathbf{E}_{L}\left( L\right) =0$ where $\mathbf{E}_{L}$ is the
Euler-Lagrange vector. Let $\bar{X}=Z^{A}\left( t,q^{B},\dot{q}^{B}\right)
\partial _{A}$ be the generator of a contact transformation. Then, if
there exist a function $F=F\left( t,q^{B},\dot{q}^{B}\right) $ such as \cite{Katzin}%
\begin{equation}
\bar{X}^{\left[ 1\right] }L=\frac{dF}{dt}  \label{LB.04c}
\end{equation}%
the vector field $X$ is a contact Noether symmetry of the Lagrangian $L\left(
t,q^{A},\dot{q}^{A}\right) ~$\cite{Sarlet}. Where $\bar{X}^{\left[ 1\right] }
$ is the first prolongation of $\bar{X}$, i.e. $\bar{X}^{\left[ 1\right] }=X+%
\dot{Z}^{i}\partial _{q^{i}}$. Furthermore, we have that for every contact
Noether symmetry there exist a function $I=I\left( t,q^{B},\dot{q}%
^{B}\right) $ \cite{Sarlet,Crampin}
\begin{equation}
I\left( t,q^{B},\dot{q}^{B}\right) =Z^{A}\left( t,q^{B},\dot{q}^{B}\right)
\frac{\partial L}{\partial \dot{q}^{A}}-F  \label{LB.04d}
\end{equation}%
such as $\frac{dI}{dt}=0$, i.e. the function $I\left( t,q^{B},\dot{q}^{B}\right)
$ is a first integral of the Lagrange equations and called a contact
Noether Integral.

In the following section we study the contact symmetries of singular
constrained Lagrangian systems.

\section{Contact transformations linear in velocities}
\label{section3}
We are seeking variational symmetries of the action
\begin{equation}  \label{act}
\mathcal{A} = \int\!\! dt L=\int\!\! dt \left(\frac{1}{2 N} G_{\mu\nu}(q)
\dot{q}^\mu \dot{q}^\nu - N\, V(q)\right), \quad \quad
\begin{matrix}
\mu,\nu= 1,...,d \\
\det{G_{\mu\nu}}\neq 0,%
\end{matrix}%
\end{equation}
generated by contact transformations, that are at most linear in the velocities $\dot{q}^\mu$. As it is well known, \eqref{act} describes a
singular system consisting of $d+1$ degrees of freedom. Its dynamics is
specified by the Euler - Lagrange equations:
\begin{subequations}\label{eul}
\begin{align}
\label{eulN}
\mathcal{H}\,&\dot{=}\,\frac{1}{2 N^2} G_{\mu\nu} \dot{q}^\mu \dot{q}^\nu+ V=0 \\
\label{eulq}
\mathcal{E}^\alpha\,&\dot{=}\, \ddot{q}^\alpha + \Gamma^\alpha_{\mu\nu} \dot{q}^\mu \dot{q}^\nu - \frac{\dot{N}}{N} \dot{q}^\alpha + N^2 G^{\mu\alpha} V_{,\mu}=0.
\end{align}
\end{subequations}

Owing to the re--parametrization invariance of the action \eqref{act},
system \eqref{eul} is singular. This fact is best understood by taking the
time derivative of \eqref{eulN} and substituting in the result the
accelerations $\ddot{q}^\mu$ from \eqref{eulq}, whence one gets an identity
(sometimes modulo \eqref{eulN}). The final conclusion is that equations %
\eqref{eul} cannot be solved for all $d+1$ dependent variables, but only for
$d$ of them; the remaining variable being freely specifiable. If this
variable is indeed selected to be a specific function of time, one can
integrate (in principle) \eqref{eulq} to obtain the remaining variables, in
which case equation \eqref{eulN} reduces to a relation between constants.

As shown in \cite{CDT2014}, in order to reveal all the symmetries that exist
by virtue of the constraint equation \eqref{eulN}, $N$ must be considered on
an equal footing with the configuration space coordinates $q^{\mu }$. Thus,
it is imperative to avoid any unnecessary gauge fixing, prior to the
derivation of the symmetries. In this regard, the canonical form of
generator $X$ of the symmetry transformation under consideration must be of
the form
\begin{equation}
X=\left( \xi ^{\kappa }(t,q,N)+K_{\;\alpha }^{\kappa }(t,q,N)\dot{q}^{\alpha
}\right) \frac{\partial }{\partial q^{\kappa }}+\Omega (t,q,N,\dot{q},\dot{N}%
)\frac{\partial }{\partial N},  \label{gen1}
\end{equation}%
incorporating $N$ as one of the variables. The infinitesimal criterion that
the symmetry generator must satisfy is condition (\ref{LB.04c}).

The first prolongation of the generator is
\begin{equation}
X^{\left[ 1\right] }=X+\phi ^{\kappa }\frac{\partial }{\partial \dot{q}%
^{\kappa }}  \label{prolong1}
\end{equation}%
with
\bal
\phi^\kappa &= \frac{d \xi}{dt}+ \frac{d}{dt} \left(K^\kappa_{\;\alpha} \dot{q}^\alpha\right)\non\\
& =  \xi^\kappa_{,t} + \xi^\kappa_{,\alpha} \dot{q}^\alpha + \xi^\kappa_{,0} \dot{N} + K^\kappa_{\;\alpha,t} \dot{q}^\alpha + K^\kappa_{\;\alpha,\beta}\dot{q}^\alpha \dot{q}^\beta
+ K^\kappa_{\; \alpha,0} \dot{N} \dot{q}^\alpha - K^\kappa_{\;\alpha} \Gamma^\alpha_{\mu\nu} \dot{q}^\mu \dot{q}^\nu+ \non \\
&\ph{=\, } \frac{1}{N} K^\kappa_{\;\alpha} \dot{N} \dot{q}^\alpha - N^2 K^{\kappa\mu} V_{,\mu} \non
\eal
where $\ph{a}_{,0}=\frac{\partial}{\partial N}$ and equations \eqref{eulq} have been used, so as to substitute the accelerations $\ddot{q}^\alpha$ with respect to the velocities $\dot{q}^\alpha$, $\dot{N}$. Note that \eqref{prolong1} needs not a prolongation term in $\frac{\partial}{\partial \dot{N}}$, since the Lagrangian is singular and thus does not involve the time derivative of the lapse $N$.

It is true that
\begin{equation} \label{cr1}
\begin{split}
\phi^\kappa \frac{\partial L}{\partial \dot{q}^\kappa} = & \frac{1}{N} \xi_{\mu , t} \dot{q}^\mu + \frac{1}{N} G_{\mu\kappa} \xi^\kappa_{,\alpha} \dot{q}^\alpha \dot{q}^\mu + \frac{1}{N} \xi_{\mu ,0} \dot{N} \dot{q}^\mu + \frac{1}{N}K_{\lambda\alpha , t}\dot{q}^\lambda\dot{q}^\alpha \\ &+ \frac{1}{N}G_{\lambda\kappa} K^\kappa_{\;\alpha,\beta}\dot{q}^\alpha\dot{q}^\beta\dot{q}^\lambda +  \frac{1}{N}K_{\lambda\alpha,0} \dot{N} \dot{q}^\alpha \dot{q}^\lambda -  \frac{1}{N}K_{\lambda\alpha} \Gamma^\alpha_{\mu\nu} \dot{q}^\mu \dot{q}^\nu \dot{q}^\lambda  \\ & +\frac{1}{N^2} K_{\lambda\alpha} \dot{N} \dot{q}^\lambda\dot{q}^\alpha - N K_\sigma^{\; \mu} V_{,\mu} \dot{q}^\sigma
\end{split}
\end{equation}
and
\be
\begin{split} \label{cr2}
\left(\xi^\kappa +K^\kappa_{\;\alpha}\dot{q}^\alpha \right) \frac{\partial L}{\partial q^\kappa} = & \frac{1}{2N} \xi^\kappa G_{\mu\nu ,\kappa} \dot{q}^\mu \dot{q}^\nu - N\xi^\kappa V_{,\kappa} + \frac{1}{2N} G_{\mu\nu,\kappa} K^\kappa_{\;\alpha} \dot{q}^\alpha \dot{q}^\mu \dot{q}^\nu \\ & - N V_{,\kappa} K^\kappa_{\; \sigma} \dot{q}^\sigma.
\end{split}
\ee
We can easily observe that only up to cubic terms in the velocities appear in the last two equations: $\dot{N} \dot{q}^2$ and $\dot{q}^3$. Hence, it is reasonable to consider adopting a specific form for $\Omega$ and $F$, so that no terms exceeding the required order are generated. However, this is not a restriction: the existence of any term not complying with this selection is trivial, since for any such addition - for example in $\Omega$ - a proper term can also be used in $F$ so that they cancel each other out. In this sense we define:
\begin{subequations}
\begin{align}
\Omega & = \omega (t,q,N) + \omega^{(1)}_\alpha (t,q,N) \dot{q}^\alpha + \omega^{(2)} (t,q,N) \dot{N} \label{omegdef}\\
F & = f (t,q,N) + f^{(1)}_\alpha (t,q,N) \dot{q}^\alpha + f^{(2)}_{\alpha\beta} (t,q,N)  \dot{q}^\alpha \dot{q}^\beta \label{fdef}
\end{align}
\end{subequations}
with $f^{(2)}_{\alpha\beta}$ being symmetric in its indices. As one can see, the gauge function $F$ has been assumed to be free of $\dot{N}$. The reasoning behind this is that no terms involving $\ddot{N}$ should be produced. This acceleration is impossible to be extracted from the derivatives of \eqref{eul}, since we have already chosen a replacement rule for the maximal number of accelerations that we are allowed to use by the equations of motion, namely the $\ddot{q}^\mu$'s. Under the above mentioned assumptions, we get
\begin{equation} \label{cr3}
\begin{split}
\Omega \frac{\partial L}{\partial N} = &- \frac{\omega}{2N^2}G_{\mu\nu}\dot{q}^\mu\dot{q}^\nu - \omega V - \frac{\omega^{(1)}_\alpha}{2N^2}G_{\mu\nu}\dot{q}^\mu\dot{q}^\nu \dot{q}^\alpha - V \omega^{(1)}_\alpha \dot{q}^\alpha \\
& - \frac{\omega^{(2)}}{2N^2}G_{\mu\nu}\dot{N}\dot{q}^\mu\dot{q}^\nu - V \omega^{(2)} \dot{N}
\end{split}
\end{equation}
and
\begin{equation} \label{cr4}
\begin{split}
\frac{dF}{dt} = &f_{,t} + f_{,\alpha}\dot{q}^\alpha + f_{,0}\dot{N} - f^{(1)}_\alpha \Gamma^\alpha_{\mu\nu} \dot{q}^\mu\dot{q}^\nu + \frac{1}{N} f^{(1)}_\alpha \dot{N} \dot{q}^\alpha - N^2 f^{(1)}_\alpha G^{\mu\alpha} V_{,\mu} \\ &+ f^{(1)}_{\alpha,t} \dot{q}^\alpha + f^{(1)}_{\alpha,\beta} \dot{q}^\alpha \dot{q}^\beta + f^{(1)}_{\alpha,0} \dot{N} \dot{q}^\alpha - 2 f^{(2)}_{\alpha\beta} \Gamma^\alpha_{\mu\nu} \dot{q}^\mu \dot{q}^\nu \dot{q}^\beta + \frac{2}{N} f^{(2)}_{\alpha\beta} \dot{N} \dot{q}^\alpha \dot{q}^\beta \\ &- 2N^2 f^{(2)}_{\alpha\beta} G^{\mu\alpha} V_{,\mu} \dot{q}^\beta + f^{(2)}_{\alpha\beta,t} \dot{q}^\alpha \dot{q}^\beta + f^{(2)}_{\alpha\beta,\gamma} \dot{q}^\alpha \dot{q}^\beta \dot{q}^\gamma + f^{(2)}_{\alpha\beta, 0} \dot{N} \dot{q}^\alpha \dot{q}^\beta
\end{split}
\end{equation}
where again \eqref{eulq} have been applied where necessary.

When relations \eqref{cr1}, \eqref{cr2}, \eqref{cr3} and \eqref{cr4} are substituted in the condition \eqref{LB.04c}, one can demand the vanishing of the coefficients appearing in front of terms involving different powers of the velocities. This is so because none of the these quantities depends on either $\dot{q}^\mu$ or $\dot{N}$. By working in this way, from the coefficients of $\dot{N}\dot{q}^2$, we get
\be \label{ena}
K_{(\mu\alpha),0} + \frac{1}{N} K_{(\mu\alpha)} - \frac{\omega^{(2)}}{2 N} G_{\mu\alpha} = 2 f^{(2)}_{\mu\alpha} + N f^{(2)}_{\mu\alpha,0}.
\ee
while out of the cubic terms $\dot{q}^3$ we derive
\be \label{dyo}
K_{(\alpha\beta ; \mu)} - \frac{1}{2N} \omega^{(1)}_{(\alpha}G_{\mu\beta)} = N f^{(2)}_{(\alpha\beta ; \mu)},
\ee
From the coefficients of $\dot{N}\dot{q}$ it follows that
\be
f^{(1)}_{\mu,0}+ \frac{1}{N} f^{(1)}_\mu = \frac{1}{N} \xi_{\mu ,0}. \non
\ee
This equation can be easily integrated to give
\be \label{tria}
\xi_\mu = N f^{(1)}_\mu + \tilde{\xi}_\mu (t,q).
\ee
Subsequently, the coefficients of the quadratic terms $\dot{q}^2$ lead to
\be \label{tessera}
\frac{1}{2} \mathcal{L}_\xi G_{\alpha\mu}+ K_{(\alpha\mu), t} - \frac{\omega}{2 N} G_{\alpha\mu} = N f^{(1)}_{(\alpha ; \mu)} + N f^{(2)}_{\alpha\mu,t}.
\ee
Finally, from the linear terms in $\dot{q}$ and $\dot{N}$, as well as the zero order terms involving no velocities we arrive at
\begin{align}
\label{pente}
\omega^{(2)} &= -\frac{1}{V} f_{,0} \\ \label{exi}
\omega^{(1)}_\mu &= \frac{1}{NV} \xi_{\mu ,t} - \frac{2 N}{V} K_{(\mu\sigma)} V^{,\sigma} - \frac{1}{V} f_{,\mu} -\frac{1}{V} f^{(1)}_{\mu,t} + \frac{2N^2}{V} f^{(2)}_{\mu\sigma} V^{,\sigma} \\ \label{epta}
\omega &= - \frac{N}{V}\xi^\kappa V_{,\kappa} -\frac{1}{V} f_{,t} + \frac{N^2}{V} f^{(1)}_\alpha V^{,\alpha}
\end{align}

We proceed in order to solve the system. Substitution of \eqref{pente} in \eqref{ena} turns the latter into
\be \label{oktw}
K_{(\mu\alpha)} = - \frac{f}{2 N V} G_{\mu\alpha} + N f^{(2)}_{\mu\alpha} + \frac{1}{N} S_{\mu\alpha}(t,q),
\ee
where $S_{\mu\alpha}$ is a symmetric matrix. By inserting \eqref{oktw} in \eqref{dyo} and with the help of \eqref{exi} we deduce that
\be
S_{(\alpha\beta ;\mu)} + \frac{V^{,\sigma}}{V} S_{\sigma (\mu}G_{\alpha\beta)}= \frac{1}{2NV} \partial_t \xi_{(\mu}G_{\alpha\beta)} - \frac{1}{2V} \partial_t f^{(1)}_{(\mu}G_{\alpha\beta)} \non,
\ee
which due to \eqref{tria} becomes
\be
S_{(\alpha\beta ;\mu)}(t,q) +\frac{V^{,\sigma}}{V} S_{\sigma (\mu}G_{\alpha\beta)}= \frac{1}{2NV} \partial_t \tilde{\xi}_{(\mu} G_{\alpha\beta)}. \non
\ee
From this equation we reach the conclusion that - since $S_{\alpha\beta}$, $G_{\alpha\beta}$, $\tilde{\xi}^\mu$ and $V$ are not functions of $N$ - the following relations
\begin{align}
\tilde{\xi}_\mu & = \tilde{\xi}_\mu(q) \\
S_{(\alpha\beta ;\mu)}& = - \frac{V^{,\sigma}}{V} S_{\sigma (\mu}G_{\alpha\beta)}
\end{align}
must hold.

At this point, we substitute \eqref{epta} in \eqref{tessera} and by virtue of \eqref{oktw} we are led to
\be
\mathcal{L}_\xi G_{\alpha\mu} +\frac{1}{N} S_{\alpha\mu ,t} + \frac{1}{2V} \xi^\kappa V_{,\kappa} G_{\alpha\mu} - \frac{N}{2V} f^{(1)}_\sigma V^{,\sigma} G_{\alpha\mu} = N f^{(1)}_{(\alpha ;\mu)}
\ee
which, by use of \eqref{tria}, becomes
\be
\frac{1}{2} \left(\mathcal{L}_{\tilde{\xi}}G_{\alpha\mu}+ \frac{V^{,\sigma}}{V} \tilde{\xi}_\sigma G_{\alpha\mu}\right) +\frac{1}{N} S_{\alpha\mu ,t}(t,q) =0\non.
\ee
Again, due to the fact that none of the remaining functions depends on $N$, we have
\begin{align}
S_{\alpha\mu} &= S_{\alpha\mu}(q) \\
\mathcal{L}_{\tilde{\xi}}G_{\alpha\mu} & =- \frac{V^{,\sigma}}{V} \tilde{\xi}_\sigma G_{\alpha\mu}.
\end{align}
Finally, we are left with the relations
\begin{subequations}
\begin{align} \label{fin1}
\xi_\mu &= N f^{(1)}_\mu (t,q,N)+ \tilde{\xi}_\mu (q) \\ \label{fin2}
K_{(\alpha\mu)} &= N f^{(2)}_{\alpha\mu}(t,q,N) - \frac{f(t,q,N)}{2NV} G_{\alpha\mu} +\frac{1}{N} S_{\alpha\mu}(q) \\
\omega &= - \frac{N}{V} \tilde{\xi}^{\sigma} V_{, \sigma} - \left(\frac{f}{V}\right)_{,t} \label{fin_omega}\\
\omega^{(1)}_\mu &= - \frac{2}{V} S_{\mu\sigma} V^{,\sigma} - \left(\frac{f}{V}\right)_{,\mu}  \label{fin_omega1}\\
\omega^{(2)} &= -\left(\frac{f}{V}\right)_{,0} \label{fin_omega2}
\end{align}
\end{subequations}
together with the conditions
\begin{subequations}  \label{sym}
\begin{align} \label{sym1}
\mathcal{L}_{\tilde{\xi}}G_{\alpha\mu} & =- \frac{V^{,\sigma}}{V} \tilde{\xi}_\sigma G_{\alpha\mu}\\ \label{sym2}
S_{(\alpha\beta ;\mu)}& = - \frac{V^{,\sigma}}{V} S_{\sigma (\mu}G_{\alpha\beta)}.
\end{align}
\end{subequations}

We recognize \eqref{sym1} as the variational Noether (conditional) symmetries found in \cite{CDT2014} for singular Lagrangians ensuing from cosmological models. These symmetries are generated by vectors $\tilde{\xi}^\mu$ in the configuration space, that are simultaneous conformal Killing vectors of the mini-supermetric $G_{\alpha\beta}$ and the potential $V$ with opposite conformal factors. The second condition, \eqref{sym2}, is the one that has to be satisfied by a symmetric tensor $S_{\alpha\beta}$ in order for a contact symmetry generator to exist. As can be seen, $S_{\alpha\beta}$ must be a conformal Killing tensor with a conformal factor that depends on the potential $V$.

At this point let us make a comparison between regular systems and re-- parametrization invariant theories (with quadratic Lagrangians): Regarding the former, we know that the relevant conditions require $\tilde{\xi}^\mu$ to be a Killing field of both the mini-supermetric and the potential \cite{Tsamparlis:Geod}, i.e.
\be
\mathcal{L}_{\tilde{\xi}}G_{\alpha\mu} =0 \quad \text{and} \quad \tilde{\xi}^\mu V_{,\mu}=0
\ee
and, in the case of linear contact symmetries, $S_{\alpha\beta}$ to be a Killing tensor \cite{Kalotas:Noether} satisfying
\be
S_{(\alpha\beta ;\mu)}=0 \quad \text{and} \quad S_{\alpha}{}^\beta V_{,\beta}=f_{,\alpha},
\ee
where $f$ is a gauge function. As one can see in each case (either Noether point or contact symmetry), two major equations have to be satisfied for regular systems. On the other hand, singular Lagrangians require one condition for each type of symmetry (\eqref{sym1} or \eqref{sym2}). This allows for an equal or (in most cases) larger group of symmetries to be admitted in the case of reparametrization invariant systems. A regular and a singular system with the same configuration space metric $G_{\alpha\beta}$ and the same potential $V$, do not necessarily exhibit the same number of symmetries; see the relevant discussion in the last section.

We can reformulate the above conditions \eqref{sym} in order to show explicitly the conformal nature of the them as follows
\bal
\mathcal{L}_{\tilde{\xi}}G_{\alpha\mu}& =\omega G_{\alpha\mu}, & \mathcal{L}_{\tilde{\xi}} V+\omega V&=0 \non\\
S_{(\alpha\beta ;\mu)}&=\psi_{(\mu}G_{\alpha\beta)},  &  S_{\sigma\mu} V^{,\sigma}+\psi_\mu V&=0. \non
\eal

Additionally, in the case of singular systems one can make use of the re--parametrization invariance of the theory to simplify conditions \eqref{sym}. This can be done in the constant potential parametrization, where the scaling $N\mapsto \xbar{N}= N\, V$ is performed. Under this change, the Lagrangian reads
\be
L= \frac{1}{2 \xbar{N}} \xbar{G}_{\mu\nu} \dot{q}^\mu \dot{q}^\nu - \xbar{N},
\ee
with $\xbar{G}_{\mu\nu}= V G_{\mu\nu}$ being the scaled, by the potential, mini-supermetric. In this parametrization, were $\xbar{V}=1$, relations \eqref{sym} reduce to
\bal  \label{symsc}
\mathcal{L}_{\tilde{\xi}}\xbar{G}_{\alpha\mu}
 =0,\quad \bar{S}_{(\alpha\beta ;\mu)}& = 0,
\eal
the covariant derivative being now constructed by $\xbar{G}_{\mu\nu}$ and the two tensors $S_{\alpha\beta},\xbar{S}_{\alpha\beta}$ related by $\xbar{S}_{\alpha\beta}=V^2\, S_{\alpha\beta}$. Thus, the symmetry generators become Killing fields and Killing tensors. All the above results can be gathered in order to formulate the following theorem:
\begin{theorem}\label{Symmetries}
For the constrained Lagrangian in action \eqref{act} and the symmetry generator \eqref{gen1} the following statements are equivalent
\begin{enumerate}
\item The vector field $\tilde{\xi}^\alpha$ is a conformal Killing vector field of the metric $G_{\alpha\beta}$ and the tensor $S_{\alpha\beta}$ is a conformal Killing tensor of the metric $G_{\alpha\beta}$ obeying the conditions
\bsub
\bal
\mathcal{L}_{\tilde{\xi}}G_{\alpha\mu}& =\omega G_{\alpha\mu}, & \mathcal{L}_{\tilde{\xi}} V+\omega V&=0 \label{killing_gen}\\
S_{(\alpha\beta ;\mu)}&=\psi_{(\mu}G_{\alpha\beta)},  &  S_{\sigma\mu} V^{,\sigma}+\psi_\mu V&=0\label{Condition}
\eal
\esub
\item The vector field $\tilde{\xi}^\alpha$ is a Killing vector field of the scaled metric $\xbar{G}_{\alpha\beta}=VG_{\alpha\beta}$ and the tensor $\xbar{S}_{\alpha\beta}$ is a Killing tensor of the conformal metric $\xbar{G}_{\alpha\beta}$
\bal
\mathcal{L}_{\tilde{\xi}}G_{\alpha\mu}=0, \quad  \xbar{S}_{(\alpha\beta ;\mu)}=0.
\eal
\end{enumerate}
\end{theorem}

\section{Conserved quantities quadratic in velocities}

\label{section4}

We now investigate the conserved quantities that are produced by the
generator we derived in the previous section. From the previous section we
have that the general form of the generator $X$ of a contact Noether symmetry for the singular Lagrangian (\ref{act})
is as follows:
\begin{equation}
\begin{split}
X& =\left( Nf^{(1)\kappa }(t,q,N)+\tilde{\xi}^{\kappa }(q)\right) \frac{%
\partial }{\partial q^{\kappa }} \\
& +\left( Nf_{\mu }^{(2)\kappa }(t,q,N)-\frac{f(t,q,N)}{2NV}\delta _{\mu
}^{\kappa }+\frac{1}{N}S_{\mu }^{\kappa }(q)\right) \dot{q}^{\mu }\frac{%
\partial }{\partial q^{\kappa }} \\
& -\left( \frac{N}{V}\tilde{\xi}^{\sigma }V_{,\sigma }+\left( \frac{f}{V}%
\right) _{,t}\right) \frac{\partial }{\partial N}-\left( \frac{2}{V}S_{\mu
\sigma }V^{,\sigma }+\left( \frac{f}{V}\right) _{,\mu }\right) \dot{q}^{\mu }%
\frac{\partial }{\partial N}-\left( \frac{f}{V}\right) _{,0}\dot{N}\frac{%
\partial }{\partial N}.
\end{split}
\notag
\end{equation}

However, as we shall immediately see, the terms involving the arbitrary
functions $f$, $f_{\mu }^{(1)}$ and $f_{\mu \nu }^{(2)}$ are trivial. From (%
\eqref{LB.04d}) we split the produced quantity $I$ into four parts
\begin{equation}
I=Q+I_{0}+I_{1}+I_{2}  \notag
\end{equation}%
with
\begin{align}
Q& =\left( \tilde{\xi}^{\alpha }+\frac{1}{N}S_{\;\mu }^{\alpha }\dot{q}^{\mu
}\right) \frac{\partial L}{\partial \dot{q}^{\alpha }}  \label{Q} \\
I_{0}& =-\frac{f}{2NV}\dot{q}^{\alpha }\frac{\partial L}{\partial \dot{q}%
^{\alpha }}  \label{I0} \\
I_{1}& =Nf_{\mu }^{(1)}G^{\mu \alpha }\frac{\partial L}{\partial \dot{q}%
^{\alpha }}  \label{I1} \\
I_{2}& =Nf_{\mu \sigma }^{(2)}G^{\mu \alpha }\dot{q}^{\sigma }\frac{\partial
L}{\partial \dot{q}^{\alpha }}.  \label{I2}
\end{align}

It is easy to prove that $I_0$, $I_1$ and $I_2$ construct trivial
integrals of motion, i.e. the corresponding conserved quantities are zero on
the constraint surface.
\begin{itemize}
\item For $I_0$:
     \be
     I_0 - f = -\frac{f}{2NV}\dot{q}^\alpha \frac{1}{N}G_{\mu\alpha}\dot{q}^\mu -f =-f \left(\frac{1}{2N^2 V} G_{\mu\nu} \dot{q}^\mu\dot{q}^\nu +1\right) =0 \non
     \ee
     due to the constraint equation \eqref{eulN}.
\item In the case of $I_1$:
     \be
     I_1 - f^{(1)}_\mu \dot{q}^\mu = N f^{(1)}_\mu G^{\mu\alpha} \frac{1}{N}G_{\sigma\alpha}\dot{q}^\sigma - f^{(1)}_\mu \dot{q}^\mu \equiv 0. \non
     \ee
\item Lastly for $I_2$:
     \be
     I_2 - f^{(2)}_{\mu\nu} \dot{q}^\mu \dot{q}^\nu = N f^{(2)}_{\sigma\mu} G^{\sigma\alpha}\dot{q}^\mu \frac{1}{N}G_{\nu\alpha}\dot{q}^\nu - f^{(2)}_{\mu\nu} \dot{q}^\mu \dot{q}^\nu \equiv 0. \non
     \ee
\end{itemize}
Thus, without any loss of generality we can consider the gauge function $F$ to be constant and deduce that the only existing conserved quantity is
\be \label{Qlin_qua}
Q = \tilde{\xi}^\alpha \frac{\partial L}{\partial \dot{q}^\alpha} + S^{\alpha\beta} \frac{\partial L}{\partial \dot{q}^\alpha} \frac{\partial L}{\partial \dot{q}^\beta},
\ee
as derived by the only non trivial part of the generator
\be
X = \left(\tilde{\xi}^\alpha + \frac{1}{N} S^\alpha_{\; \mu}\dot{q}^\mu\right) \frac{\partial}{\partial \dot{q}^\alpha} - \frac{V^{,\mu}}{V}\left(N \tilde{\xi}_\mu+ 2S_{\mu\nu}\dot{q}^\nu\right)\non \frac{\partial}{\partial N},
\ee
with $\tilde{\xi}^\alpha$ and $S_{\alpha\beta}$ satisfying \eqref{sym1} and \eqref{sym2} respectively. As seen by \eqref{Qlin_qua},
in the phase space, the vector $\tilde{\xi}^\alpha$ gives rise to
integrals of motion linear in the momenta $p_\alpha=\frac{\partial L}{\partial \dot{q}^\alpha}$,
while the tensor $S^{\alpha\beta}$ quadratic. As a result, we can state the following:
\begin{theorem}\label{Integrals}
A singular system described by a Lagrangian of the general form
\be \label{lagth}
L=\frac{1}{2N} G_{\mu\nu}(q) \dot{q}^\mu \dot{q}^\nu - N \, V(q),
\ee
admits an integral of motion quadratic in the momenta of the form ($p_\mu = \frac{\partial L}{\partial \dot{q}^\mu}$)
\be
Q = S^{\alpha\beta}(q)\, p_\alpha p_\beta,
\ee
when $S_{\alpha\beta}$ obeys condition \eqref{Condition}.
\end{theorem}

Conformal Killing tensors (CKTs) can be divided into two major classes: reducible and irreducible. The first type consists of tensors that are trivially constructed as tensor products of other CKTs of lower rank. In the case of second rank conformal Killing tensors, the reducible CKTs are made up by conformal Killing vectors. Of course, if there exists an integral of motion linear in the momenta, we expect its square to be again a constant of motion, i.e. if there exist $\tilde{\xi}^\mu$'s that satisfy \eqref{sym1}, second rank tensor products can be constructed by them satisfying \eqref{sym2}. Indeed, let us consider a second rank tensor $S_{\alpha\beta}=\Lambda^{IJ} \tilde{\xi}_I^\alpha \tilde{\xi}_J^\beta$, with $\Lambda^{IJ}$ being a symmetric constant matrix and $\tilde{\xi}_I^\mu$, $\tilde{\xi}_J^\mu$ vectors that satisfy \eqref{sym1} (the indexes $I$, $J$ are just used to discriminate among the various vectors). Under these considerations we write
\begin{equation*}
\begin{split}
S_{(\alpha\beta; \mu)} = &\Lambda^{IJ} (\tilde{\xi}_{I(\alpha} \tilde{\xi}_{J\beta})_{;\mu)} = \Lambda^{IJ} \left(\tilde{\xi}_{I(\alpha ;\mu} \tilde{\xi}_{J\beta)} + \tilde{\xi}_{I(\alpha} \tilde{\xi}_{J\beta ;\mu)} \right)  \\
=& \frac{\Lambda^{IJ}}{6} \left[ \left(\tilde{\xi}_{I\mu ;\beta} + \tilde{\xi}_{I\beta;\mu}\right) \tilde{\xi}_{J\alpha} +\left(\tilde{\xi}_{I\alpha ;\mu} + \tilde{\xi}_{I\mu;\alpha}\right) \tilde{\xi}_{J\beta} + \left(\tilde{\xi}_{I\alpha ;\beta} + \tilde{\xi}_{I\beta;\alpha}\right) \tilde{\xi}_{J\mu} \right. \\
 & \left. +
\tilde{\xi}_{I\alpha}\left(\tilde{\xi}_{J\beta ;\mu} + \tilde{\xi}_{J\mu;\beta}\right)  + \tilde{\xi}_{I\beta} \left(\tilde{\xi}_{J\alpha ;\mu} + \tilde{\xi}_{J\mu;\alpha}\right)  + \tilde{\xi}_{I\mu} \left(\tilde{\xi}_{J\alpha ;\beta} + \tilde{\xi}_{J\beta;\alpha}\right) \right] \\
=& -\frac{\Lambda^{IJ}}{6} \frac{V^{,\sigma}}{V} \left( \tilde{\xi}_{I\sigma}\tilde{\xi}_{J\beta} G_{\alpha\mu} + \tilde{\xi}_{I\sigma}\tilde{\xi}_{J\alpha} G_{\beta\mu} + \tilde{\xi}_{I\sigma}\tilde{\xi}_{J\mu} G_{\alpha\beta} \right. \\
& \left. + \tilde{\xi}_{I\alpha}\tilde{\xi}_{J\sigma} G_{\beta\mu} + \tilde{\xi}_{I\beta}\tilde{\xi}_{J\sigma} G_{\alpha\mu} + \tilde{\xi}_{I\mu}\tilde{\xi}_{J\sigma} G_{\alpha\beta} \right) \\
=& - \Lambda^{IJ} \frac{V^{,\sigma}}{V} \tilde{\xi}_{I\sigma} \tilde{\xi}_{J(\mu}G_{\alpha\beta)} \\
= & - \frac{V^{,\sigma}}{V} S_{\sigma(\mu}G_{\alpha\beta)}
\end{split}
\end{equation*}
and thus symmetries defined by $\tilde{\xi}^\alpha$'s can be trivially used to construct contact symmetries generated with the help of reducible conformal Killing tensors. An occurrence quite common in cosmology is that of conformally flat configuration
metrics in mini-superspace; in this event it is well known that irreducible CKTs do exist, see \cite{Rani:killing}.

\section{Examples}\label{examples}
In order to make things more transparent we apply the previously developed general theory to two specific examples; the first is the determination of the geodesics of a pp--wave spacetime while the second concerns the constrained two dimensional Hamiltonian Ermakov--Ray--Reid system.

\subsection{Determination of Geodesics}

When the potential $V(q)$ in the action \eqref{act} is constant the corresponding Lagrangian can be used in order to calculate the geodesics of a manifold with metric tensor $G_{\mu\nu}$; equation \eqref{eulq} represents the geodesic equation in the non-affine parameter $t$. Since we have theorem \eqref{Integrals} at hand we can fix the gauge, i.e. choose a particular functional form for $N(t)$; the most natural choice for the time variable is $N(t)=1$ which gives the time variable $t=s$ the role of an \emph{affine parameter}.

In \cite{Keane2010} the authors calculated the Killing tensors for a number of pp--wave spacetimes along with their conformal Killing vector fields. The metric tensor $G$ for the type $Biv$ pp--wave spacetime admits the form
\bal\label{Biv_metric}
G=G_{\alpha\beta} \drm q^\alpha \otimes \drm q^\beta=-2\,\drm u\otimes \drm v-\frac{2}{z^2}\,\drm u \otimes \drm u+\drm y \otimes \drm y+\drm z \otimes \drm z.
\eal

The conformal algebra of the above spacetime is six dimensionsal ${\cal S}_6 \supset {\cal H}_5 \supset {\cal G}_4$ (four Killing vectors, one homothetic, and one conformal) with basis
\bal
X_1 &= \partial_v, & X_2 &= \partial_u, \qquad X_3 = \partial_y, \qquad X_4 = y \partial_v + u \partial_y
\nonumber\\
X_5 &= 2u \partial_u + y \partial_y + z \partial_z, &
X_6 &= u^2 \partial_u + \tfrac{1}{2} (y^2 + z^2) \partial_v + u (y \partial_y + z \partial_z).
\nonumber
\eal
The five irreducible Killling tensors are
\bal
(S_1)_{\alpha\beta} &= -2 y^2 z^{-2} \delta_{(\alpha}^u \delta_{\beta)}^u - z^2 \delta_{(\alpha}^y \delta_{\beta)}^y
+ 2 yz \delta_{(\alpha}^y \delta_{\beta)}^z - y^2 \delta_{(\alpha}^z \delta_{\beta)}^z
\nonumber\\
(S_2)_{\alpha\beta} &= 2yz^{-2} \delta_{(\alpha}^u \delta_{\beta)}^u - z \delta_{(\alpha}^y \delta_{\beta)}^z
+ y \delta_{(\alpha}^z \delta_{\beta)}^z
\nonumber\\
(S_3)_{\alpha\beta} &= 2 u yz^{-2} \delta_{(\alpha}^u \delta_{\beta)}^u + z^2 \delta_{(\alpha}^u \delta_{\beta)}^y
-yz \delta_{(\alpha}^u \delta_{\beta)}^z - uz \delta_{(\alpha}^y \delta_{\beta)}^z + uy \delta_{(\alpha}^z \delta_{\beta)}^z
\nonumber\\
(S_4)_{\alpha\beta} &= 2 u z^{-2} \delta_{(\alpha}^u \delta_{\beta)}^u - z \delta_{(\alpha}^u \delta_{\beta)}^z
+ u \delta_{(\alpha}^z \delta_{\beta)}^z
\nonumber\\
(S_5)_{\alpha\beta} &= (z^2 + 2u^2 z^{-2}) \delta_{(\alpha}^u \delta_{\beta)}^u
- 2uz \delta_{(\alpha}^u \delta_{\beta)}^z + u^2 \delta_{(\alpha}^z \delta_{\beta)}^z.
\nonumber
\eal

Since in this case the potential is zero the symmetry generators are the four Killing vectors $X_i,\,i=\xbar{1,4}$ and the five Killing tensors $S_j,\,j=\xbar{1,5}$, which produce the nine integrals of motion $\{I_i=(X_i)_\alpha p^\alpha,\, Q_j=(S_j)_{\alpha\beta} p^\alpha p^\beta\}$ along with the quadratic constraint $\mathcal{H}$ \eqref{eulN}:
\bsub
\bal
I_1&=-\dot{u}, \quad I_2=-2z^{-2}\dot{u}-\dot{v}, \quad I_3=\dot{y}, \quad I_4=\dot{y}u-y\dot{u} \\
Q_1&=z^{-2}\left( - z^4\dot{y}^2+2y z^3 \dot{y} \dot{z}-\left(2\dot{u}^2+z^2\dot{z}^2 \right)y^2 \right) \\
Q_2&=z^{-2} \left( 2y\dot{u}^2-z^3 \dot{y} \dot{z}+y z^2 \dot{z}^2  \right)\\
Q_3&=z^{-2}\left( z^3\dot{u}\left( z\dot{y}-\dot{z}y \right)+
u y\left(2\dot{u}^2+z^2\dot{z}^2 \right)-u z^3\dot{y} \dot{z}  \right)\\
Q_4&=z^{-2} \left( 2 u \dot{u}^2-z^3 \dot{u} \dot{z}+u  z^2 \dot{z}^2 \right)\\
Q_5& = z^{-2}\left( u^2\left( 2\dot{u}^2+z^2 \dot{z}^2\right)+ z^4 \dot{u}^2-2u z^3 \dot{u} \dot{z}  \right)\\
\mathcal{H}&=\tfrac{1}{2}z^{-2} \left(  z^2 \left( 2-2\dot{u} \dot{v}+\dot{y}^2+\dot{z}^2 \right)-2\dot{u}^2 \right),\label{ham}
\eal
\esub
where the dot represents differentiation with respect to the affine parameter $s$. From the integrals $I_1,I_3$ we can solve for $u(s),y(s)$ i.e.
\bal
u(s)=-I_1\,s+c_u,\, y(s)=I_3\,s+c_y,
\eal
where $c_u,c_y$ are constants of integration. The next step is to eliminate the $\dot{z}^2$ from the integrals $Q_2,Q_4$ using a linear combination of them, yielding
\bal
\frac{dz^2}{ds}=J_1+2J_2 s\Rightarrow z^2=J_2 s^2+J_1 s+c_z, \non
\eal
where $c_z$ is another constant of integration and the constants $J_1,J_2$ are combinations of the existing constants, i.e. $J_1=2\left(Q_4 c_y-Q_2 c_u  \right)/\left( c_y I_1+c_u I_3 \right),J_2=\left(Q_2 I_1+Q_4 I_3  \right)/\left( c_y I_1+c_u I_3 \right)$. Finally we are left with the function $v(s)$ which can be determinated from the integral $I_2$ which now assumes the form
\bal
\dot{v}=\frac{2I_1}{J_2 s^2+J_1 s+cz}-I_2\Rightarrow v(s)=\frac{4I_1}{\sqrt{4c_z J_2-J_1^2}}\arctan \frac{2J_2 s+J_1}{\sqrt{4c_z J_2-J_1^2}}-I_2 s+c_v,\non
\eal
with $c_v$ the last constant of integration. If one counts the constants that appear in the solution space the total number of them is $9$ but the actual number that describes the geodesics equation is $2\times 4 = 8$, thus there must exist a relation between the $9$ constants. The desired relation comes from the integral $Q_2$ and reads $c_z=\left( 8 I_1^2+J_1^2  \right)/(4 J_2)$.

Gathering the above results the final form of the solution space is
\bal\label{sol_geo}
u(s)&=-I_1 s+c_u, \quad &v(s)&=\sqrt{2} \arctan\frac{2J_2 s+J_1}{2\sqrt{2} I_1}-I_2 s+c_v \non\\
y(s)&=I_3 s+c_y, \quad & z^2(s)&=\frac{\left( 2J_2 s+J_1 \right)^2+8 I_1^2}{4 J_2}.
\eal

It is interesting to note that, since we have not used the Hamiltonian constraint \eqref{ham} to solve the system, the solution \eqref{sol_geo} describes an \emph{unconstrained} system; if someone insists on the validity of the constraint, then as stated before, a relation among the constants emerges, i.e.
\bal
J_2+I_3^2-2 I_1 I_2+2=0.\non
\eal

\subsection{Hamiltonian Ermakov--Ray--Reid system.}

The two dimensional constraint Hamiltonian Ermakov--Ray--Reid system can be described by the Hamiltonian
\bal\label{2d Ermakov}
H=\frac{1}{2N}\left( p_\alpha^2+p_\beta^2 \right)+N V,\quad V=\frac{1}{2}\omega(\alpha^2+\beta^2)+\frac{1}{\alpha^2} J\left(\frac{\beta}{\alpha}\right),
\eal
with $\omega,J$ arbitrary functions of their arguments. The origins of this system can be traced back in 1880 in the pioneer work of Ermakov \cite{Ermakov} (see \cite{Harin:Ermakov} for the English translation) along with the generalization of Ray and Reid \cite{Ray_Reid,Reid_Ray}; for the unconstrained Hamiltonian version see  \cite{Rogers:Optics,Rogers:pulsrodon}. The theoretical interest in this system resides in its admittance of an integral of motion for every function $\omega,J$, namely, the Ray-Reid invariant
\bal\label{RR:inv}
I=\frac{1}{2}\left( \dot{\alpha}\beta-\alpha\dot{\beta} \right)^2+\left( \frac{\alpha^2+\beta^2}{\alpha^2} \right) J\left(
\frac{\beta}{\alpha} \right).
\eal

The above integral can be produced from a Lie point symmetry with the help of Noether's theorem, as first pointed out by Lutzky in \cite{Lutzky:Integral}. Here we are going to show how the integral \eqref{RR:inv} emerges naturally from the contact symmetries of the Lagrangian.

The metric tensor $G_{\mu\nu}=\delta_{\mu\nu}$ emerging from \eqref{2d Ermakov} corresponds to a flat two--dimensional space. It is well known that the Killing and/or conformal Killing tensors span an infinite dimensional space when the dimension of the base manifold space equals to two. The CKTs $S_{\alpha\beta}$ along with the corresponding conformal factors $\psi_\alpha$ are
\bal\label{2d CKT}
S_{\alpha\beta}=\begin{pmatrix}
F(\alpha,\beta) & k_1-\tfrac{\ima}{2}\left( f_1(z)-f_2(\bar{z})  \right) \\
k_1-\tfrac{\ima}{2}\left( f_1(z)-f_2(\bar{z})  \right) &
F(\alpha,\beta)+f_1(z)+f_2(\bar{z})
\end{pmatrix},
\eal
and
\bal\label{2d factor}
\psi_\alpha=\left( \partial_\alpha F(\alpha,\beta),\partial_\beta F(\alpha,\beta)+f_1'(z)+f_2'(\bar{z}) \right),
\eal
where $z=\beta+\ima\alpha,\bar{z}=\beta-\ima\alpha$ and $k_1=const.$

In order to apply theorem \eqref{Symmetries} we must demand the validity of the condition \eqref{Condition} where the potential and the CKTs are given by equations \eqref{2d Ermakov} and \eqref{2d CKT} respectively. It is interesting that we can satisfy condition \eqref{Condition} for every function $\omega$ and $J$ when the functions $F,f_1,f_2$ are given by
\bal\non
F(\alpha,\beta)=\frac{\left( 4c_2 J(\beta/\alpha)-2 c_2 \beta^2 \omega(\alpha^2+\beta^2)+c_3 \right)\alpha^2}{2 J(\beta/\alpha)+\alpha^2 \omega(\alpha^2+\beta^2)}\\
f_1(z)=c_1+c_2 z^2,\, f_2(z)=-c_1+c_2 z^2\non
\eal

With the above choice the resulting CKTs are
\bal\label{CKT_Ermakov}
(S_1)_{\alpha\beta} &=
\begin{pmatrix}
\dfrac{2 \alpha^2 J(w)-\alpha^2\beta^2 \omega(u)}{2J(w)+\alpha^2 \omega(u)} & \alpha\beta\\
\alpha\beta & \dfrac{2\beta^2 J(w)-\alpha^4 \omega(u)}{2J(w)+\alpha^2 \omega(u)}
\end{pmatrix}\\
(S_2)_{\alpha\beta} &=
\begin{pmatrix}
\dfrac{\alpha^2}{2J(w)+\alpha^2 \omega(u)} & 0\\
0 & \dfrac{\alpha^2}{2J(w)+\alpha^2 \omega(u)}
\end{pmatrix},
\eal
where $w=\beta/\alpha,u=\alpha^2+\beta^2$. From these CKT, the respective integrals of motion along with the Hamiltonian constraint read
\bsub
\bal
Q_1&=\frac{2\left(\alpha \dot{\alpha}+\beta \dot{\beta} \right)^2 J(w)- \alpha^2 \left(  \dot{\alpha}\beta-\alpha\dot{\beta} \right)^2 \omega(u)}{2J(w)+\alpha^2 \omega(u)} \\
Q_2&=\frac{\left( \dot{\alpha}^2+\dot{\beta}^2 \right) \alpha^2}{2J(w)+\alpha^2 \omega(u)}\\
\mathcal{H}&=\frac{1}{2}\left( \dot{\alpha}^2+\dot{\beta}^2 \right)+ \frac{1}{2}\omega(u)+\frac{1}{\alpha^2} J(w).
\eal
\esub
Solving the Hamiltonian constraint  for the function $\omega(u)$ and substituting in $Q_1$ we recover
the Ray--Reid invariant \eqref{RR:inv}.
We would like to remark that with a similar approach we can construct and the correspoding conservation law of the
generalized Ermakov system in an $n$-dimensional Riemannian space \cite{MTermakov}.

\section{Discussion}
\label{disc}

In this work we have generalized previous results in \cite{CDT2014} concerning the determination of Lie point -- Noether symmetries  for constraint systems whose action is quadratic in the velocities. The generalization concerns the consideration of contact symmetries, i.e. of generators which depend linearly on the velocities.

The key ingredients of the present approach are:
\begin{enumerate}[label=(\alph*)]
\item the treatment of the lapse function $N(t)$ and $q^\alpha(t)$'s on an equal footing in the generator \eqref{gen1}, despite the fact that the corresponding Euler--Lagrange equation is $\partial_N L\equiv 0$, which reveals a larger number of symmetries and
\item the dependence of the gauge function $F(t,q^\alpha,\dot{q}^\alpha)$ \eqref{fdef} in the velocities $\dot{q}^\alpha$ which in turn makes it constant, i.e. non--essential in contrast to the regular case.
\end{enumerate}

If someone chooses not to implement the $N$--dependence in generator \eqref{gen1} then the function $\Omega(t,q^\alpha,\dot{q}^\alpha)$ \eqref{omegdef} should be zero, i.e. $\omega=0,\omega^{(1)}_\mu=0,\omega^{(2)}=0$. Then equation \eqref{fin_omega1} $\omega=0$ yields $f=f(q^\alpha),\mathcal{L}_{\tilde{\xi}}V=0$ and equation \eqref{killing_gen} reads $\mathcal{L}_{\tilde{\xi}}G_{\mu\nu}=0$. Furthermore from equation \eqref{fin_omega1} $\omega^{(1)}_\mu=0$ we have the equality $S_{\mu\sigma}V^{,\sigma}=-V/2\left( f/V  \right)_{,\mu}$ which forces the conformal factor $\psi_\mu$ to be $\psi_\mu=\frac{1}{2}\left( f/V \right)_{,\mu}$ due to equation \eqref{Condition}. At this stage if we redefine the conformal Killing tensor $S_{\alpha\beta}$ via $S_{\alpha\beta}=\tilde{S}_{\alpha\beta}+\frac{1}{2}\left( f/V \right)G_{\mu\nu}$ then we arrive at $\tilde{S}_{\alpha\beta}=0$. Recapitulating the above situation we conclude that if we do not use the $N$--dependence the symmetry generators are described by the \emph{Killing vector fields} and \emph{Killing tensor fields} of the metric $G_{\mu\nu}$ and not the \emph{conformal} ones. Thus if we had at hand a metric $G_{\mu\nu}$ that does not admits Killing fields but do admits conformal ones then one could conclude that there are no symmetries, which is of course a fault result. In order to give an example let us introduce the $2d$ metric $g$:
\bal\label{special metric}
g=g_{\alpha\beta}\drm x^\alpha\otimes \drm x^\beta\Rightarrow g=\frac{x^2y^2-1}{y^2}\left( x\drm x\otimes \drm x +y\drm y\otimes \drm y\right).
\eal
It is an easy task for one to see that metric \eqref{special metric} does not admit Killing vectors but since it is two dimensional admits an infinite number of conformal Killing fields. If we choose the potential $V=\left( y(x^2y^2-1) \right)^{-1}$ then there exist the two conformal Killing fields
\bal
\xi_1=\frac{1}{\sqrt{x}}\partial_x,\quad \xi_2=x\partial_x+y\partial_y
\eal
that satisfy condition \eqref{killing_gen}, thus generating two symmetries. Equivalently one can state that \emph{fixing the gauge might result in loosing symmetries}.

As far as the issue concerning the dependence of the gauge
function $F$ in the velocities $\dot{q}^\alpha$, we would like to stress the
following facts: Let's for a moment follow the common practice and demand
no velocity dependence in $F$, i.e. $F=F(t,q^\alpha)$, repeating
the procedure for the evaluation of the symmetries we arrive instead of
condition \eqref{Condition} to the condition
\bal
 S_{\sigma\mu}V^{,\sigma}+\psi_\mu V+\frac{1}{2}f_{,\mu}=0, \non
\eal
which might tempt one to think that it is more general, due to the appearance
of the arbitrary function $f(q^\alpha)$; however, this is not the case:
If we redefine the conformal Killing tensor $S_{\alpha\beta}$ in the same
spirit as we did before; i.e. $S_{\mu\nu}=\widetilde{S}_{\mu\nu}-\frac{1}{2}\left( f/V \right)G_{\mu\nu}$ the
above equation reduces to condition \eqref{Condition} signaling the non--essentiality of the gauge function $f$.
As we have shown the function $f$ generates the Hamiltonian constraint $\mathcal{H}$, see equation \eqref{I0},
thus one should have expected the non--necessity of the gauge function $f$, due to the presence of the natural
"gauge function" of the problem in hand, i.e. the constraint $\mathcal{H}$.

Furthermore, our results can also be used to the unconstraint case,
provided that we have enough symmetries to use, so that
we do not need the constraint $\mathcal{H}$; as, for example, it happens
the case of section \ref{examples}, where we calculated
the geodesics of the pp--wave spacetime.

Finally, we conclude that the determination of the Noether symmetries (point and contact alike) of systems with Lagrangian
\eqref{act}, has been reduced to a problem of differential geometry; that is, to the determination of the
Killing vectors and Killing Tensors of conformally related spacetimes.

\section*{Acknowledgements}
ND acknowledges financial support by FONDECYT postdoctoral grant no. 3150016.
AP acknowledges financial support of INFN.

\phantomsection
\addcontentsline{toc}{section}{References}

\bibliographystyle{utphys}
\bibliography{Main}

\end{document}